\documentclass[table,xcdraw,runningheads]{llncs}
\usepackage{graphicx}
\usepackage{subcaption} 
\usepackage[linesnumbered, ruled,vlined]{algorithm2e}
\usepackage[draft]{changes}
\usepackage[bookmarks=false]{hyperref}
\usepackage{enumitem}
\usepackage{tabularx}
\usepackage{wrapfig}
\usepackage{wasysym}
 \usepackage{booktabs}
\usepackage[table,xcdraw]{}

\let\llncssubparagraph\subparagraph
\let\subparagraph\paragraph
\usepackage[compact]{titlesec}
\let\subparagraph\llncssubparagraph

\definecolor{random}{rgb}{0.55, 0.45, 0.39}
\definecolor{darkpastelred}{rgb}{0.76, 0.23, 0.13}
\definecolor{darkviolet}{rgb}{0.58, 0.0, 0.83}
\definechangesauthor[name=Enes Erdin, color=darkviolet]{EE}
\definechangesauthor[name=Mumin Cebe, color=random]{MC}

\begin{document}
\title{LNBot: A Covert Hybrid Botnet on Bitcoin Lightning Network for Fun and Profit}
%
%
\author{Ahmet Kurt \and
Enes Erdin \and
Mumin Cebe \and
Kemal Akkaya \and \\
A. Selcuk Uluagac}
\authorrunning{A. Kurt et al.}
%
\institute{Florida International University, Miami FL 33174, USA\\
\email{\{akurt005,eerdi001,mcebe,kakkaya,suluagac\}@fiu.edu}}
\maketitle              
\begin{abstract}
While various covert botnets were proposed in the past, they still lack complete anonymization for their servers/botmasters or suffer from slow communication between the botmaster and the bots. In this paper, we propose a new generation hybrid botnet that covertly and efficiently communicates over Bitcoin Lightning Network (LN), called LNBot. LN is a payment channel network operating on top of Bitcoin network for faster Bitcoin transactions with negligible fees. Exploiting various anonymity features of LN, we designed a scalable two-layer botnet which completely anonymize the identity of the botmaster. In the first layer, the botmaster sends commands anonymously to the C\&C servers through LN transactions. Specifically, LNBot allows botmaster's commands to be sent in the form of surreptitious multihop LN payments, where the commands are encoded with ASCII or Huffman encoding to provide covert communications. In the second layer, C\&C servers further relay those commands to the bots they control in their mini-botnets to launch any type of attacks to victim machines. We implemented a proof-of-concept on the actual LN and extensively analyzed the delay and cost performance of LNBot. Our analysis show that LNBot achieves better scalibility compared to the other similar blockchain botnets with negligible costs. Finally, we also provide and discuss a list of potential countermeasures to detect LNBot activities and minimize its impacts. 


\end{abstract}
%
%
%
\section{Introduction}
Botnets are networks of computing devices infected with malicious software that is under the control of an attacker, known as bot herder or \textit{botmaster} \cite{SILVA2013378}. The owner of the botnet controls the \textit{bots} (i.e., devices that become part of the botnet) through command and control \textit{(C\&C) server(s)} which can communicate with the bots using a C\&C channel and can launch various attacks through these bots, including, but not limited to, denial of service (DoS) attacks, information and identity theft, sending spam messages, and other activities. Naturally, a botmaster's goal is to make it difficult for law enforcement to detect and prevent malicious operations. Therefore, establishing a secure C\&C infrastructure and hiding the C\&C server identity play a key role in the long-lasting operation of botnets. 

Numerous botnets have been proposed and deployed in the past \cite{Grizzard,Wang}. Regardless of their communication infrastructure being centralized or peer-to-peer, existing botnet C\&C channels and servers have the challenge of remaining hidden and being resistant against legal authorities' actions. Such problems motivate hackers to always explore more promising venues for finding new C\&C channels with the ever-increasing number of communication options on the Internet. One such platform is the environment where cryptocurrencies, such as Bitcoin, is exchanged. As Bitcoin offers some degree of anonymity, exploiting it as a C\&C communication channel has already been tried for creating new botnets~\cite{ali2018zombiecoin,ali2015zombiecoin}. While these Bitcoin-based botnets addressed the long transaction validation times, they still announce the commands publicly, where the botnet activity can be traced by any observer with the help of the Bitcoin addresses or nonce values of the transactions. By using Bitcoin for botnet communications, C\&C leaves the history of malicious activities on the blockchain forever.

Nonetheless, the issues regarding the public announcement of commands and leaving traces in the blockchain are already being addressed in a newly developed Bitcoin payment channel network called Lightning Network (LN). LN enables off-line transactions (i.e., transactions which are not announced and thus not recorded on the blockchain) in order to speed up the transaction by eliminating the confirmation time and decreasing fees associated with that transaction. Additionally, users' identities are still kept anonymous since the transactions are not announced publicly. In this paper, we advocate using LN as an ideal C\&C infrastructure for botnets with all the aforementioned features (i.e., faster transactions, decreased costs). Specifically, LN offers botmasters numerous advantages over existing techniques: First, LN provides very high anonymity since transactions on the off-chain channels are not recorded on the blockchain. Thus, a botmaster can covertly communicate with the C\&C server(s). Second, the revelation of a server does not reveal other servers, and an observer cannot enumerate the size of the botnet. Most importantly, C\&C communication over the LN cannot be censored. 

Although LN is a fast-growing emerging payment network, it only has around 12K nodes which may not be ideal for large-scale botnets. Therefore, we propose a \textit{two-layer hybrid} botnet to use LN as an infrastructure to maintain a network of C\&C servers each of which can run its own botnet. The use of multiple C\&C servers has been around for a while \cite{ollmann2009botnet}. However, the communication with these servers was still assumed to be through the existing communication infrastructures which impairs the servers' anonymity.
Therefore, further strengthening of anonymity is still needed.

Hence, this paper presents \textit{LNBot}, which is the first botnet that utilizes LN infrastructure for its communications between the botmaster and C\&C servers with a two-layer \textit{hybrid architecture}. Specifically, at the first layer, a botmaster 
will maintain multiple C\&C servers,  which are nodes on the LN that have specialized software to control the bots under them. Essentially, each C\&C server is controlling an independent isolated mini-botnet at the second layer. These mini-botnets are controlled using a specific C\&C infrastructure that can rely on 
traditional means such as stenography, IRC channel, DNS, Tor, etc. Botmaster sends the commands to the C\&C servers covertly through LN. This two-layer command and control topology not only enables scalability, but also minimizes the burden on each  C\&C server, which increases their anonymity.

To demonstrate the feasibility of the concept, we implemented the \textit{LNBot} using real LN nodes in Bitcoin's Testnet which is the actual network for Bitcoin. Utilizing one-to-many architecture (i.e., botmaster sends commands to all C\&C servers separately), we show that by encoding the commands in terms of payments sent over LN, one can successfully send commands to the C\&C servers that are part of the LN. These C\&C servers further relay those commands to the bots they control to launch any type of attack to victim machines. 

Nevertheless, as sending the commands to every C\&C server in the form of payment requires the botmaster to maintain high capacity LN channels (i.e., increased monetary cost) and pay forwarding fees to LN, we also propose mechanisms to further decrease these costs to the levels where they can be neglected. Specifically, when the attacks are executed, we circulate the received payments at C\&C servers back to botmaster. Essentially, this means the botmaster will receive all of his/her money back except the fees charged by LN. To also minimize those fees, in addition to ASCII-based encoding, we propose a Huffman-based encoding mechanism which considers the frequency of characters that could potentially be used in constructing the attack commands. We demonstrate that for a network comprising 100 C\&C servers, the total fixed fees for forming LNBot would be lower than \$5. 



Contrary to the traditional blockchain based communication schemes, LNBot covertly communicates with the C\&C servers by utilizing the strong relationship anonymity of LN. This covert communication comes with a very little cost and latency overhead. Additionally, since LNBot does not require a custom C\&C infrastructure, it is very practical to deploy it. All these features of LNBot makes it a botnet that needs to be taken seriously therefore we provide a list of countermeasures that may help detect LNBot activity and minimize damages from it.

The rest of the paper is organized as follows: In Section \ref{sec:background}, we give some background information about LN. In Section \ref{sec:LNBot}, we describe the architecture and construction of our proposed LNBot. Section \ref{sec:ProofOfConcept} is dedicated to proof-of-concept implementation in real world settings while Section \ref{sec:evaluation} presents the evaluation results. In Section \ref{sec:countermeasures}, possible countermeasures for LNBot is discussed. Related work is given in Section \ref{sec:relatedwork}. Finally, we conclude the paper in Section \ref{sec:conclusion}.

\section{Background}
\label{sec:background}


\subsection{Lightning Network}

The LN concept is introduced in \cite{poon2015bitcoin}. It is a payment protocol operating on top of Bitcoin. Through this concept, an overlay payment network (i.e., LN) is started among the customers and vendors in 2017. 
The aim in creating the LN was to decrease the load on the Bitcoin network, facilitating transactions with affordable fees and reduced transaction validation times, and increasing the scalability of Bitcoin by establishing peer-to-peer connections. Despite the big fluctuations in the price of Bitcoin recently, the LN grew exponentially reaching 12,384 nodes and 36,378 channels in less than two years by the time of writing this paper \cite{LN-size}. In the following subsections, we briefly explain the components of LN. 

\subsection{Off-chain Concept}
\label{sec:offchainconcept}

\begin{figure}[htb]
\vspace{-6mm}
  \centering
  \includegraphics[width=\linewidth]{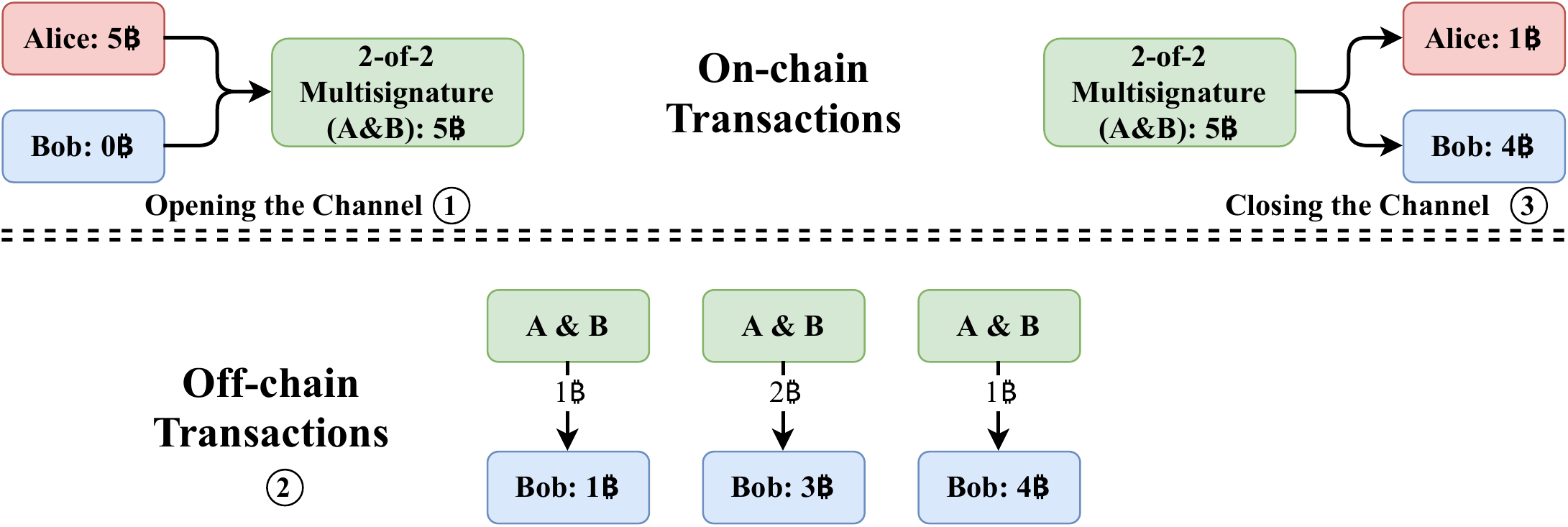}
     \vspace{-2mm}
  \caption{Off-chain mechanism of LN.}
  \label{fig:offchain}
   \vspace{-7mm}
\end{figure}

The main idea behind LN is to utilize the \textit{off-chain} concept \cite{pass2015micropayments} which enables near-instant Bitcoin transactions with negligible fees. This is accomplished through \textit{bidirectional payment channels} which are created among two parties to exchange funds without committing the transactions to Bitcoin blockchain. The channel is opened by making a single on-chain transaction called \textit{funding transaction}. 
The funding transaction determines the capacity of the channel. Whenever one of the parties wants to make a transaction, she basically conveys ownership of that amount of her money to her peer. So, after a transaction takes place the total capacity in the channel does not change but the directional capacities do. Therefore, the peers can make as many as transactions they want in any amount unless the amount they want to send does not exceed the directional capacity.
The example shown in Fig. \ref{fig:offchain} illustrates the concept in more detail.

Per figure, \textcircled{{\scriptsize \textbf{1}}} Alice opens a channel to Bob by funding 5 Bitcoins to a multi-signature address and the multi-signature address is signed by both Alice and Bob.  \textcircled{{\scriptsize \textbf{2}}} Using this channel, Alice can send payments to Bob simply by transferring the ownership of her share in the multi-signature address until her fund in the  address is exhausted. Note that these transactions are off-chain meaning they are not written to the Bitcoin blockchain which is a unique feature of LN and that feature is exploited in our botnet. Alice performs 3 transactions at different times with amounts of 1, 2 and 1 Bitcoin respectively.  \textcircled{{\scriptsize \textbf{3}}} Eventually, when the channel is closed, the remaining 1 Bitcoin in the multi-signature wallet is returned to Alice while the total transferred 4 Bitcoins are settled at Bob. Channel closing is another on-chain transaction that reports the final balances of Alice and Bob in the multi-signature address to the blockchain. 

\subsection{Multihop Payments}

\begin{figure}[htb]
\vspace{-5.5mm}
  \centering
  \includegraphics[width=0.53\textwidth]{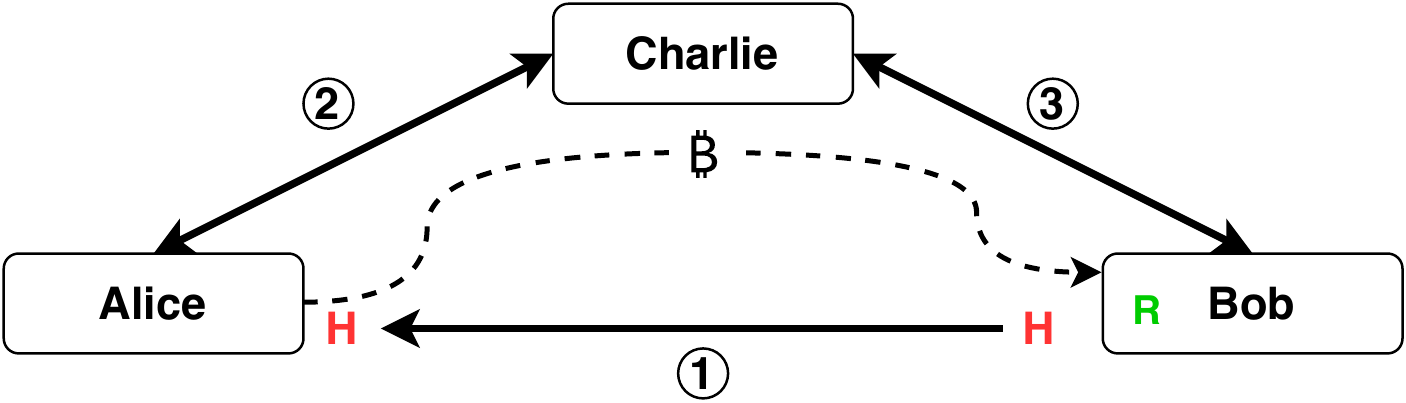}
  \vspace{-1mm}
  \caption{Illustration of a multihop payment. $R$ is the secret (i.e. \textit{pre-hash}) generated by Bob, $H$ is the hash of the secret. When the transaction is locked by $H$, yielding the secret $R$ opens the lock. Namely, when asked, yielding $R$ changes the ownership of the money in the channel.}
  \vspace{-4mm}
  \label{fig:multihop}
\end{figure}

In LN, a node can make payments to another node even if it does not have a direct payment channel to that node. This is achieved by utilizing already established payment channels between other nodes and such a payment is called multi-hop payment since the payment is forwarded through a number of intermediary nodes until reaching the destination. This process is trustless meaning the sender does not need to trust the intermediary nodes along the route. Fig. \ref{fig:multihop} depicts a multi-hop payment. As there is a direct payment channel between Alice and Charlie and between Charlie and Bob, Alice can initiate a transaction to Bob via Charlie. \textcircled{{\scriptsize \textbf{1}}} First, Bob sends an invoice to Alice which includes the hash ($H$) of a secret $R$ (known as \textit{pre-image}). \textcircled{{\scriptsize \textbf{2}}} Then, Alice prepares a payment locked with $H$, the answer of which is known by Bob. Hash-Locking is required for Alice to ensure that the payment is received by Bob. So, locked with $H$, Alice gives ownership of some of her money destined to Bob if and only if Charlie knows and discloses the answer to $H$. Likewise, \textcircled{{\scriptsize \textbf{3}}} Charlie promises to give the ownership of some of his money which is locked by $H$ to Bob if Bob knows the answer. As Bob receives a payment he naturally discloses the answer to Charlie and in return he gets the money from Charlie as promised. Now, as Charlie learned the answer, he discloses the answer to Alice and gets his money from Alice as promised. This mechanism is realized with the ``Hash Time Lock Contracts'' (HTLC). Through this mechanism of LN, as long as there is a path from source to destination requiring that the channels on the path have enough capacities, payments can be routed just like the Internet.

 \vspace{-3mm}



\subsection{\textit{Key Send} Payments}
\label{sec:sphinx}


\textit{Key send} in LN enables sending payments to a destination without needing to have an invoice first \cite{sphinxsend}. It utilizes \textit{Sphinx}\cite{Sphinx} which is a compact and secure cryptographic packet format to anonymously route a message from a sender to a receiver. This is a very useful feature to have in LN because it introduces new use cases where payers can send spontaneous payments without contacting the payee first. In this mode, the sender generates the pre-image for the payment instead of the receiver and embeds the pre-image into the sphinx packet within the outgoing HTLC. If an LN node accepts key send payments, then it only needs to advertise its public key to receive key send payments from other nodes. In LNBot, we utilize this feature to send payments from botmaster to C\&C servers.

\subsection{Source Routing \& Onion Routed Payments}
\label{sec:onionrouting}
With the availability of multi-hop payments, a routing mechanism is needed to select the best route for sending a payment to its destination. LN utilizes \textit{source routing} which gives the sender full control over the route for the transaction to follow within the network. Senders are able to specify: 1) the total number of hops on the path; 2) total cumulative fee they will pay to send the payment; and 3) total time-lock period for the HTLC 
\cite{onionrouting2}. Moreover, all payments in LN are \textit{onion-routed payments} meaning that HTLCs are securely and privately routed within the network. Onion routing LN also prevents any nodes from easily censoring payments since they are not aware of the final destination of the payment. Additionally, by encoding payment routes within a Sphinx packet, following security and privacy features are achieved: the nodes on a routing path do not know: 1) the source and the destination of the payment; 2) their exact position within the route; and 3) the total number of hops in the route. 

\subsection{Motivation to Use LN for a Botnet}

In this section, we justify why LN is suitable for a perfect botnet design. 



\begin{itemize}[leftmargin=*]
    \item \textbf{No publicly advertised activity:} The drawback of using a cryptocurrency based communication infrastructure is that all of the activities are publicly stored in a persistent, transparent, append-only ledger. However, using the off-chain transaction mechanism, only the intermediary nodes in a multi-hop payment path know the transactions. Because, they are responsible to keep the state of their own channels just to prove the ownership of their share in case of a dispute. Namely, the activities taking place in a payment channel is locally stored by the nodes responsible for forwarding that transaction.
    
    \item \textbf{Covert messaging:} LN was proposed to ease the problems occurring in the Bitcoin network. Hence, all of the actions taking place in the network is regarded as financial transactions. In that sense, twisting this idea into using the channels for covertly forwarding commands will be indistinguishable from innocent, legitimate, and real financial transactions. 
    
    \item \textbf{Relationship anonymity:} LN utilizes source-routing for payment forwarding. This feature enables the peers to enjoy a higher anonymity. Assume that, during a transaction let the next node successfully guess that the preceding node was the origin of the transaction. Ideally, there is no possibility for it to successfully guess the final destination for that transaction. This applies to any ``curious'' node in the network. Namely, without collusion it is impossible to know who communicates to whom, which is known as relationship anonymity feature.
    
    \item \textbf{Instantaneous communication:} Apart from being public, another drawback of using Bitcoin network is that a transaction is approved in 10 minutes the earliest. Moreover, for a transaction to be approved in the ledger for good, the peers have to wait for at least 60 minutes. By moving to the off-chain, a transaction simply becomes a network package traversing in the network through the intermediary nodes. In that sense, the communication latency in LN is nothing but the time for a packet to traverse on the Internet.
    
    \item \textbf{Minimal cost:} Bitcoin network charges fees for every transaction regardless of its amount. LN was also designed to transform these fees into negligible amounts. 
    The fees charged by LN is comprised of the combination of a ``base fee'' and a ``proportional fee'' which are close to zero. In the \texttt{lnd} implementation of LN, default setting for the base fee is 1 millisatoshi\footnote{A satoshi is defined to be 0.00000001 Bitcoin. In other words, 1 Bitcoin is 100 million satoshi.}. The proportional fee is, as name suggests it is proportional with the amount being forwarded, 0.0001\% of the payment. 

\end{itemize}

\section{LNBot Architecture}
\label{sec:LNBot}

In this section, we describe the overall architecture of LNBot with its elements.

\subsection{Overview}
\label{sec:Overview}

\begin{figure}[htb]
    \vspace{-6mm}
    \centering
    \includegraphics[width=\linewidth]{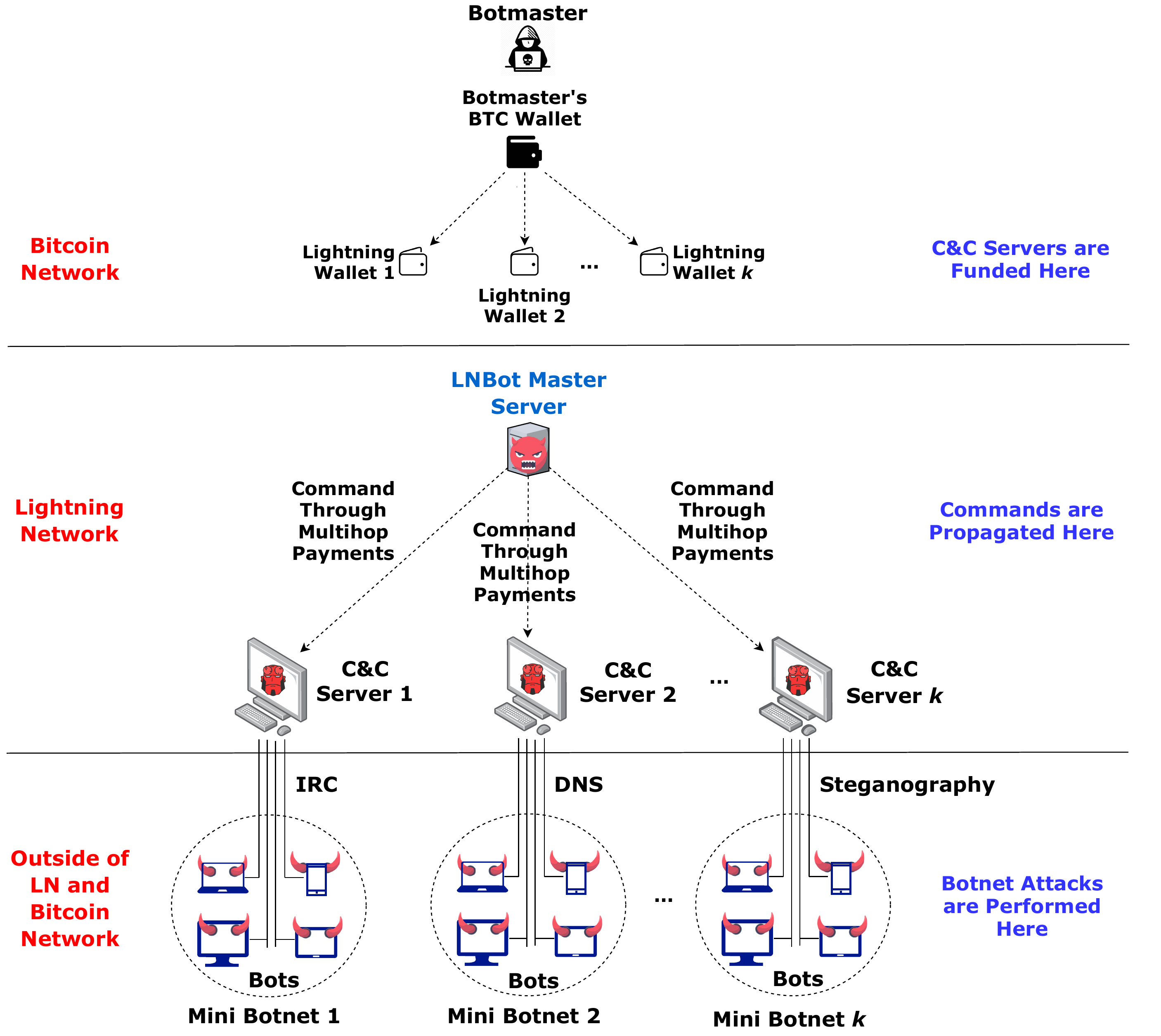}
    \vspace{-5mm}
    \caption{Overview of LNBot Architecture.}
    \label{fig:lnbotoverview}
\end{figure}

The overall architecture is shown in Fig. \ref{fig:lnbotoverview}. As shown, the LN is used to maintain the C\&C servers and their communication with the botmaster. Each C\&C server runs a separate mini-botnet. Note that it is up to the botmaster how to populate these mini-botnets. Each C\&C server can utilize a different botnet model (i.e., based on IRC, DNS, steganography, cryptocurrencies, etc). 

The botmaster could set up the C\&C servers by creating LN nodes at remote locations that are accessible to him/her. The botmaster knows the LN public keys of all C\&C servers since s/he sets them up. These public keys are needed to communicate with them in the LN. Then s/he installs a special software on the C\&C servers which are used to control the bots. In this way, it is enough for botmaster to release a malware into the wild for infecting user machines and upon infection, these machines connect to existing available C\&C servers (i.e., they become bots). One possible way to achieve this would be to spread the malware via embedded advertisements on web pages frequently visited by intended victims. When a viewer clicks on the link, s/he is redirected to a website hosting malicious code that executes in the background and infects the victim's machine without his/her knowledge. 

Upon infection, the bot establishes a communication with an available C\&C server. The type of connection used depends on the communication method chosen by the C\&C server the bot connects to. This can be picked among existing botnet C\&C infrastructures such as IRC, DNS, steganography, cryptocurrencies or even the LN itself.

The botmaster's commands have to propagate to every C\&C server, and then, ultimately to every single bot through the C\&C servers. For this task, we propose one-to-many propagation where the botmaster sends commands to each C\&C server separately. This approach is described in Section \ref{sec:onetomany}. The botmaster periodically issues commands to C\&C servers by sending payments over LN. Thus, the commands have to be encoded into a series of LN payments. We implemented two encoding schemes to represent the commands as LN payments. These methods are detailed in Section \ref{sec:encodings}.

With the availability of command propagation, the C\&C servers could now listen to the incoming instructions from the botmaster. Next, we describe the details of setting up the C\&C servers. 

\subsection{Setting up the C\&C Servers}
As mentioned earlier, the botmaster can set up the necessary number of C\&C servers s/he would like to deploy. Depending on the objectives, the number of these servers and the number of bots they will control can be adjusted without any scalability concern. In Section \ref{sec:ProofOfConcept}, we explain how we set up real C\&C servers running on LN on the real Bitcoin network.

Each C\&C server is deployed as a \textit{private} LN node which means that they do not advertise themselves within the LN. In this way, the other LN nodes do not know about the existence of the C\&C servers and cannot open channels to them without knowing their public keys. 
However, without opening any channels on the network, C\&C servers cannot get botmaster's payments on LN. Therefore, each C\&C servers open a channel to at least $k$ different random public LN nodes. To open the channels, they need some Bitcoin in their \textit{lightning wallets}. This Bitcoin is provided to C\&C servers by the botmaster before deploying them. The number $k$ may be tuned depending on the size and topology of LN when LNBot will have deployed in the future.



\subsection{Formation of Mini-botnets}
After C\&C servers are set up, we need bots to establish connections to C\&C servers. An infected machine (bot) connects to one of the C\&C servers.
As mentioned earlier, the details of bot recruitment and any malware implementation issues are beyond the objectives of this paper. It is up to the botmaster to decide which type of infrastructure the C\&C servers will use to control the bots in their possession. This flexibility is enabled by our proposed two-layer hybrid architecture of LNBot. The reason for giving this flexibility is to enable scalability of LNBot through any type of mini-botnets without bothering for the compromise of any C\&C servers. 
As it will be shown in Section \ref{sec:countermeasures}, even if the C\&C servers are compromised, this neither reveals the other C\&C servers nor the botmaster.

\subsection{Forming LNBot}
Now that C\&C servers are set up and mini-botnets are formed, the next step is to form the infrastructure to control these C\&C servers covertly with minimal chances of getting detected. This is where LN comes into play. Botmaster has the public keys of all LN nodes running on C\&C server machines. Since C\&C servers have their LN channels ready, they can receive the commands from the botmaster. The botmaster uses an LN node called \textit{LNBot Master Server} to initiate the commands to all the C\&C servers through LN payments. 
Similar to the C\&C servers, LNBot Master Server is also a private LN node and botmaster has flexibility on the setup of this node and may change it regularly.
Without using any other custom infrastructure, the botmaster is able to control C\&C servers through LN, consequently controlling all the bots on the botnet.

\subsection{Command Propagation in LNBot}
\label{sec:onetomany}

Once the LNBot is formed, the next step is to ensure communication from the botmaster to the C\&C servers. We utilize a \textit{one-to-many} architecture where the botmaster sends the commands to each C\&C server separately. The botmaster uses \textit{key send} method mentioned in Section \ref{sec:sphinx} to send the payments. 
We designed a command sending protocol for botmaster-to-C\&C server communication as shown in Algorithm 1.

\SetKw{KwTo}{in}
\begin{algorithm}[htb]
\label{algo:algorithm1}
\SetAlgoLined
initialize $command$\;
int $counter$ = 0\;
bool $isOnline$ = checkIfC\&CServerIsOnline()\;
\uIf {$isOnline$}{
    bool $result$ = send($5$ $satoshi$)\;
    \uIf {$result$=success}{
        $counter$ = 0\;
        \For{$character$ \KwTo $command$}{
            bool $result$ = send($character$)\;
            \lIf {$result$=success}{continue}
            \uElseIf {$result$=fail and $counter<k$} 
                {retry sending $character$\;
                $counter$++;}
            \lElse {reschedule($command$, date, time)}
        }
        $counter$ = 0\;
        bool $result$ = send($6$ $satoshi$)\;
        \uIf {$result$=success}
            {\textbf{Command has been successfully sent!};}
        \uElseIf{$result$=fail and $counter<k$}
            {retry sending $6$ $satoshi$\;
            $counter$++;}
        \lElse {reschedule($command$, date, time)}
    }
    \uElseIf{$result$=fail and $counter<k$}
        {retry sending $5$ $satoshi$\;
        $counter$++;}
    \lElse{reschedule($command$, date, time)}
}
\Else{reschedule($command$, date, time);}
\caption{Send Command}
\end{algorithm}

Before sending any payment, the botmaster first checks if the respective C\&C server is online or not (LN nodes have to be online in order to send and receive payments). If the C\&C server is not online, command sending is scheduled for a later time. Botmaster sends 5 satoshi as the special starting payment of a command before it sends the actual characters in the command one by one. Lastly, the botmaster sends 6 satoshi as the special ending payment to finish sending the command. Note that selection of 5 and 6 in this algorithm depends on the chosen encoding and could be changed based on the needs. If any of these separate payments fail, it is re-tried. If any of the payments fail for more than $k$ times in a row, command transmission to the respective C\&C server is canceled and scheduled for a later time. The details of encoding and decoding are explained next. 

\subsection{Encoding/Decoding Schemes}
\label{sec:encodings}
An important feature of LNBot is its ability to encode botmaster commands into a series of LN payments. We used two different encoding/decoding schemes for the purpose of determining the most efficient way of sending commands to C\&C servers in terms of Bitcoin cost and time spent. We explain the details of each method below: 

\subsubsection{ASCII Encoding}

American Standard Code for Information Interchange (ASCII) is a character encoding standard that represents English characters as numbers, assigned from 0 to 127.  Numbers 0-31 and 127 are reserved for control characters. The remaining 95 codes from 32 to 126 represent printable characters. The decimal equivalent of ASCII characters can easily be looked up from an ASCII table.

\subsubsection{Huffman Coding}

When there is a need to losslessly compress the information being sent over a channel, due to its simple yet powerful approach, Huffman coding is one of the optimal options \cite{huffman1952method}. In usual communication systems, the communication is done in binary domain. However, in the communication scheme defined as in our approach, there is no strict need for binary communication. In the formation of the Huffman tree, $n-$ary number systems can be used. The advantage of $n-$ary numbering system over binary one is that the messages can be distributed among more compact symbols, hence the required number of transmissions per character will be reduced. 

In order to come up with a codebook, a dictionary is needed. The frequencies, so-called probabilities of occurrences of the characters shape the size of the codebook. In its most frequently adapted style, users prefer to use bulky novels or texts in order to simulate a more inclusive dictionary. 

\subsection{Reimbursing the Botmaster}


Another important feature of LNBot is the ability of the botmaster to get the invested funds back from C\&C servers' lightning wallets to his/her Bitcoin wallet. Depending on botmaster's command propagation activity, C\&C servers' channels will fill up with funds received from the botmaster. Therefore, in our design, C\&C servers are programmed to send the funds in their channels to an LN node called \textit{collector}. Collector is set up by the botmaster as a private LN node which becomes active only when the C\&C servers will send funds to it. Its LN public key is stored in C\&C servers and thus they can send the funds to collector through LN using the collector's public key. In this way, the funds accumulate at the collector. The botmaster gets the funds accumulated at the collector when his/her channels starts running out of funds. Botmaster get the funds from collector by closing collector's channels so that the funds at these channels are settled at collector's lightning wallet. Then botmaster sends these funds through an on-chain Bitcoin transfer to his/her Bitcoin wallet.

\section{Proof-of-Concept Implementation}
\label{sec:ProofOfConcept}

In this section, we demonstrate that an actual implementation of the proposed LNBot is feasible by presenting a proof-of-concept. For development, we used \texttt{lnd} (version 0.9.0-beta) which is one of the implementations of LN developed by Lightning Labs~\cite{LN}. LN nodes should interact with a Bitcoin network in order to run the underlying layer-1 protocols. There are two real environments where Bitcoin operations take place: \textit{Bitcoin Mainnet} and \textit{Bitcoin Testnet}. As the names suggest, Bitcoin \textit{Mainnet} is the chain where Bitcoin transfers with a real monetary value take place. However, in Bitcoin \textit{Testnet}, Bitcoins do not have a monetary value. They are only used for testing and development purposes. Nonetheless, they both provide the same infrastructure and LNBot will definitely run in the same manner on the \textit{Mainnet} as it runs on the \textit{Testnet}. 

Thus, we used Bitcoin \textit{Testnet} for our proof-of-concept development. We created 100 C\&C servers and assessed certain performance characteristics for command propagation. We created a GitHub page explaining the steps to set up the C\&C servers.\footnote{\url{https://github.com/LightningNetworkBot/LNBot}} 
The steps include installation of \textit{lnd} \& \textit{bitcoind}, configuring \textit{lnd} and \textit{bitcoind}, and extra configurations to hide the servers in the network by utilizing private channels. 
Nevertheless, to confirm that the channel opening costs and routing fees are exactly same in both Bitcoin \textit{Mainnet} and \textit{Testnet}, we also created 2 nodes on Bitcoin \textit{Mainnet}. We funded one of the nodes with 0.01 Bitcoin ($\sim$\$67), created channels and sent payments to the other node. We observed that the costs and fees are exactly matching to that of Bitcoin \textit{Testnet}. 


\textit{lnd} has a feature called \textit{autopilot} which opens channels in an automated manner based on certain initial parameters set in advance \cite{samplelndconf}. Our C\&C servers on Bitcoin \textit{Testnet} employ this functionality of \textit{lnd} to open channels on LN. 
Using \textit{autopilot}, we opened 3 channels per server. Note that this number of channels is picked based on our experimentation on Bitcoin \textit{Testnet} on the success of payments. We wanted to prevent any failures in payments by tuning this parameter.  As mentioned, these 3 channels are all private, created with \textit{--private} argument, which do not broadcast themselves to the network. A private channel in LN is only known by the two peers of the channel.

\textit{lnd} has an API for communicating with a local \textit{lnd} instance through gRPC~\cite{gRPC}. Using the API, we wrote a client that communicated with \textit{lnd} in Python. Particularly, we wrote 2 Python scripts, one running on the C\&C servers and the other on the botmaster machine. We typed the command we wanted to send to C\&C servers in a terminal in the botmaster machine. The command was processed by the Python code and sent to the C\&C servers as a series of payments. 

\section{Evaluation and Analysis of LNBot}
\label{sec:evaluation}

In this section, we present a detailed cost and time overhead analysis of LNBot.



\subsection{Cost Analysis of LNBot Formation}
\label{sec:LNBotcreation}

We first analyze the monetary cost of forming LNBot. 
As noted earlier, we opened 3 channels per server. The capacity of each channel is 20,000 satoshi which is the minimum allowable channel capacity in \textit{lnd}. Therefore, a server needs 60,000 satoshi for opening these channels.
While opening the channels, there is a small fee paid to Bitcoin miners since channel creations in LN are on-chain transactions. We showed that, opening a channel in LN can cost as low as 154 satoshi on both Bitcoin Testnet\footnote{Check LNB6's channel (\href{https://1ml.com/testnet/channel/1735152493945290752}{1735152493945290752}) opening transaction for instance: \newline \href{https://blockstream.info/testnet/tx/fc46c99233389d24c4fd9517cd503f08265c517a6f0570d806e7cc98b7f7963b}{fc46c99233389d24c4fd9517cd503f08265c517a6f0570d806e7cc98b7f7963b}} and the Mainnet.\footnote{In a similar way, check one of our mainnet node's channel opening transaction: \newline \href{https://blockstream.info/tx/1d81b6022ff1472939c4db730ca01b82d43b616e757d799aea17ee0db6427520}{1d81b6022ff1472939c4db730ca01b82d43b616e757d799aea17ee0db6427520}}

So the total cost of opening 3 channels for a C\&C server is 60,462 satoshi. While 462 satoshi is consumed as fees, the remaining 60,000 satoshi on the channels is not spent, rather it is just locked in the channels. The botmaster will get this 60,000 satoshi back after closing the channels. Therefore, funds locked in the channels are non-recurring investment cost for the formation of LNBot. Only real associated cost of forming LNBot is the channel opening fees.

Table \ref{tab:channelopening} shows how the costs change when the number of C\&C servers is increased. The increase in the cost is linear and for 100 C\&C servers, the on-chain fees is only 0.000462 Bitcoin (\$3 at current Bitcoin price of \$6700).


\begin{table}[htb]
\vspace{-10mm}
  \begin{center}
    \caption{Channel Opening Fees for Different Number of C\&C Servers}
    \label{tab:channelopening}
    \resizebox{0.65\linewidth}{!}{
    \begin{tabular}{|c|c|c|c|}
    \hline
      \textbf{Number of C\&C Servers} & \textbf{Channel Opening Fees}  \\
      \hline
      10 & 0.0000462 Bitcoin \\
      \hline
      25 & 0.0001155 Bitcoin \\
      \hline
      50  & 0.000231 Bitcoin \\
      \hline
      100 & 0.000462 Bitcoin\\
      \hline
    \end{tabular}
  }
   \vspace{-10mm}
  \end{center}
\end{table}



\subsection{Cost and Time Analysis of Command Propagation}
\label{sec:LNBotPropagationResults}

To assess the command propagation overhead, we sent the following SYN flooding attack command to C\&C servers from the botmaster (omitting start and end of command characters):

\texttt{sudo hping3 -i u1 -S -p 80 -c 10 192.168.1.1} \\

We sent this command using both of the encoding methods we proposed earlier. For Huffman coding, we compared several different base number systems. The best result was obtained by using the Quaternary numeral system, the codebook of which is shown in Table \ref{tab:quaternaryhuffman}.

\begin{table}[htb]
\vspace{-20pt}
\parbox{.3\linewidth}{
    \centering
    \caption{Obtained codebook for Huffman coding}
    \label{tab:quaternaryhuffman}
    \resizebox{1.1\linewidth}{!}{
    \begin{tabular}{|c c|c c|c c|c c|}
    \hline 
    `s' & 234 & `n' & 233 & `o' & 232 & `h' & 231 \\ \hline 
    `d' & 224 & `g' & 223 & `c' & 222 & `9' & 221 \\ \hline 
    `6' & 214 & `2' & 213 & `3' & 212 & `u' & 211 \\ \hline 
    `p' & 144 & `i' & 143 & `8' & 142 & `0' & 141 \\ \hline 
    `.' & 24  & `1' & 12  & `-' & 13  & `E' & 4   \\ \hline 
    ` ' & 11  & `S' & 3   &     &     &     &     \\ \hline
    \end{tabular}}
}
\hfill
\parbox{.65\linewidth}{
    \centering
    \caption{Respective ASCII and Huffman encoding representation of `sudo hping3 -i u1 -S -p 80 -c 10 192.168.1.1' command}
    \label{tab:ASCII}
    \resizebox{\linewidth}{!}{
    \begin{tabular}{|c|c|c|}
    \hline 
    \textbf{Command} & \textbf{ASCII Encoding}        & \textbf{Quaternary}     \\ 
    ~                & ~                              & \textbf{Huffman Encoding}     \\ \hline
    `sudo '          & 115,117,100,111,32             & 2,3,4,2,1,1,2,2,4,2,3,2,1,1   \\ \hline 
    `hping3 '        & 104,112,105,                   & 2,3,1,1,4,4,1,4,3,2,3,3     \\ 
     ~               & 110,103,51,32                  & 2,2,3,2,1,2,1,1             \\ \hline
    `-i '            & 45,105,32                      & 1,3,1,4,3,1,1               \\ \hline
    `u1 '            & 117,49,32                      & 2,1,1,1,2,1,1               \\ \hline
    `-S '            & 45,83,32                       & 1,3,3,1,1                   \\ \hline
    `-p '            & 45,112,32                      & 1,3,1,4,4,1,1               \\ \hline
    `80 '            & 56,48,32                       & 1,4,2,1,4,1,1,1             \\ \hline
    `-c '            & 45,99,32                       & 1,3,2,2,2,1,1               \\ \hline
    `10 '            & 49,48,32                       & 1,2,1,4,1,1,1               \\ \hline
     `192.168.1.1'   & 49,57,50,46,49                 & 1,2,2,2,1,2,1,3,2,4,1,2,2    \\ 
    ~                & 54,56,46,49,46,49              & 1,4,1,4,2,2,4,1,2,2,4,1,2    \\ \hline \hline
    Total Number of  & 44                             & 108                         \\
    Payments         & ~                              & ~                           \\ \hline
    Total Cost       & 2813                           & 215                         \\ \hline
    \end{tabular}}
}
\vspace{-6mm}
\end{table}

\noindent \textbf{Cost Analysis:} The botmaster spent 2813 satoshi for sending the SYN flooding command using the ASCII encoding while this cost is only 215 satoshi with the Huffman coding. Table \ref{tab:ASCII} gives details about the number of payments and how many satoshi have been sent in each payment. While in both cost cases the botmaster will be reimbursed at the very end, we would like to note that the lifetime of the channels is closely related with these costs. In case of the ASCII encoding, the initial funds will be spent faster and the botmaster needs to re-configure (or re-balance) the channels for continuous operation of the botnet. In case of the Huffman coding, this is not the case as the consumption of the channel funds is much slower. So, we can see that if channel lifetime is an important factor for the botmaster, the Huffman coding could be preferred. In other words, the Huffman coding gives the botmaster the ability to perform more attacks without creating high capacity channels.


However, the situation is reverse in case of routing fees. Table \ref{tab:routingfees} shows how the routing fees change when the number of C\&C servers is increased. The increase in the routing fees is linear for both the ASCII and Huffman coding. For 100 C\&C servers, total routing fee paid is only 0.000176 Bitcoin ($\sim$ \$1 at current Bitcoin price of \$6700) for ASCII while it is 0.000432 Bitcoin ($\sim$ \$3 at current Bitcoin price of \$6700) for the Huffman coding. This indicates that despite its increased routing fees, the Huffman coding is still a viable option for longer operation of LNBot. 


\vspace{-5mm}

\begin{table}[htb]
\vspace{-6mm}
  \begin{center}
    \caption{Routing Fees for Different Number of C\&C Servers}
    \label{tab:routingfees}
    \resizebox{\linewidth}{!}{
    \begin{tabular}{|c|c|c|c|}
    \hline
      \textbf{Number of C\&C Servers} & \textbf{Routing Fees (ASCII)} & \textbf{Routing Fees (Huffman)}  \\
      \hline
      10 & 0.0000176 Bitcoin & 0.0000432 Bitcoin \\
      \hline
      25 & 0.000044 Bitcoin & 0.000108 Bitcoin \\
      \hline
      50 & 0.000088 Bitcoin & 0.000216 Bitcoin \\
      \hline
      100 & 0.000176 Bitcoin & 0.000432 Bitcoin\\
      \hline
    \end{tabular}
  }
   \vspace{-10mm}
  \end{center}
\end{table}

\begin{wrapfigure}{r}{0.5\textwidth}
\vspace{-10mm}
    \includegraphics[width=0.5\textwidth]{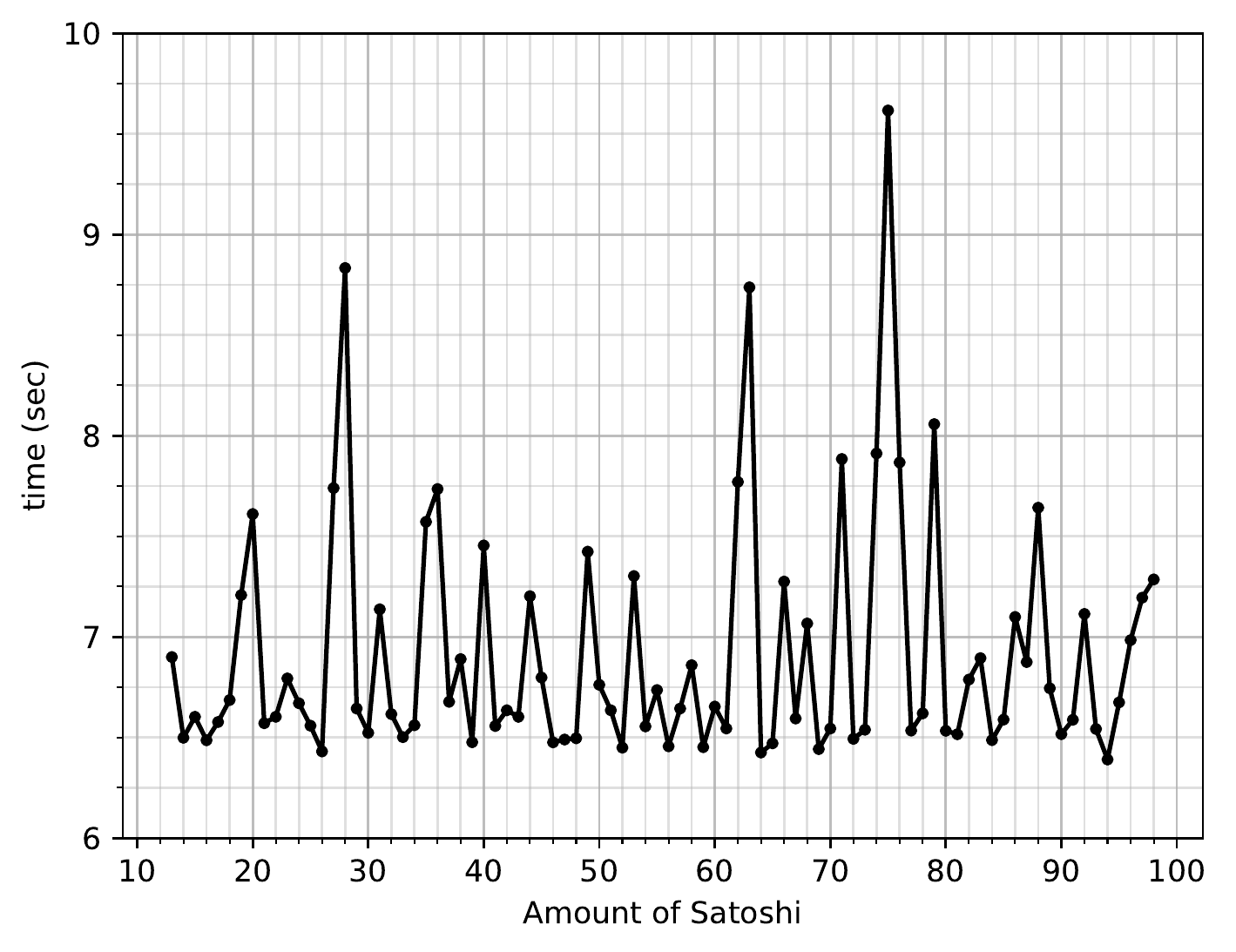}
    \vspace{-8.5mm}
    \caption{Time for \textit{key send payments} to reach their destinations with varying satoshi.}
    \label{fig:paymenttime}
    \vspace{-7.5mm}

\end{wrapfigure}

\noindent \textbf{Time Analysis:} The propagation time of a command is calculated by multiplying the number of payments with the average delivery time of the payments. To estimate the average delivery time, we sent 90 \textit{key send payments} with different amounts from botmaster to our C\&C servers over LN at random times and measured the time it takes for payments to reach their destinations. The results are depicted in Fig. \ref{fig:paymenttime}. 

As shown, \textit{key send payments} took 7 seconds on average to reach their destinations and the maximum delay was never exceeding 10 seconds. This delay varies since it depends on the path being used and the load of each intermediary node in the LN. We observed that the number of hops for the payments was 4, which helps to strengthen unlikability of payments and destinations in case of any payment analysis in LN. 

Using an average of 7 seconds, the total propagation time for the ASCII-encoded payments is 7x44=308 seconds while it is 7x108=756 seconds for the Huffman coding. The Huffman coding reduces the cost of sending the command, but increases the communication delays which is not critical in performing the attack.

\subsection{Comparison of LNBot with Other Similar Botnets}

We also considered other existing botnets that utilize Bitcoin for their command and control. Using our SYN flooding attack command, we computed their cost and command propagation times to compare them with LNBot. We also included their scalability features. Table \ref{tab:comparison} shows these results. 

\begin{table}[htp]
\vspace{-20pt}
  \begin{center}
    \caption{Time, Cost and Scalability Comparison of LNBot with Similar Botnets.}
    \label{tab:comparison}
    \resizebox{\linewidth}{!}{
    \begin{tabular}{|c|c|c|c|}
    \hline
      \textbf{Botnet} & \textbf{Cost} & \textbf{Time} & \textbf{Scalability}\\
      \hline
      \textbf{Bitcoin Testnet Botnet \cite{dazaleveraging}} & 51349 satoshi (Testnet) & $\sim$ 10 minutes & Low, thousands of bots\\
      \hline
      \textbf{Zombiecoin 2.0 \cite{ali2018zombiecoin}} & 10000 satoshi & $\sim$ 10 seconds & Low, thousands of bots \\
      \hline
      \textbf{LNBot} & 10 satoshi & $\sim$ 5 minutes & High, millions of bots\\
      \hline
    \end{tabular}
  }
  \end{center}
  \vspace{-10mm}
\end{table}

As seen, LNBot comes with minimal costs with a reasonable propagation time for attacks and can scale to millions of nodes with its two-layer architecture.

\section{Security \& Anonymity Analysis and Countermeasures} 
\label{sec:countermeasures} 
In this section, we discuss security properties of LNBot and possible countermeasures to detect its activities in order to minimize its impacts. \\




\noindent $\bullet$ \textit{Taking LN down:} Obviously, the simplest way to eliminate LNBot's activities is taking down the LN as a whole once there is any suspicion about a botnet. However, this is very unlikely due to LN being a very resilient decentralized payment channel network. In addition, today many applications are running on LN and shutting down may cause a lot of financial loss for numerous stakeholders.  

\noindent $\bullet$ \textit{Compromising and shutting down a C\&C Server:} In LNBot there are many C\&C servers each of which is controlling a mini-botnet. Given the past experience with various traditional botnets, it is highly likely that these mini-botnets will be detected at some point in the future paving the way for also the detection of a C\&C server. This will then result in the revelation of its location/IP address and eventually physical seizure of the machine by law enforcement. 
Nevertheless, the seizure of a C\&C server will neither reveal the identity of the LNBot botmaster nor other C\&C servers since a C\&C server receives the commands through onion routed payments catered with Sphinx's secure packet format, which does not reveal the original sender of the message. Additionally, the communication between botmaster and C\&C servers is 1-way meaning that botmaster can talk to C\&C servers, but servers cannot talk back since the LN address of the botmaster is not known by them. This 1-way communication ensures that the identity of the botmaster will be kept secret at all times. 

Note that since the C\&C servers hold the LN public key of the collector, it will also be revealed when a C\&C server is compromised. However, since the collector node's channels are all private, its IP address or location is not known by the C\&C servers. Therefore, learning the LN public key of the collector node does not help locating the collector node physically. The only possibility is to continue monitoring a C\&C server when it is compromised and as soon as it makes a payment (to collector), we can try to do  a timing analysis on certain random nodes that are under our control to determine if one of them would be forwarding the same amount of money and happens to have a channel with the collector node. In that case, that node will know the IP address of the collector since they have a channel. While this possibility is very low, even if we are successful, the collector can always hide its IP address through certain mechanisms such as VPN or Tor.  Eventually, we can see that taking down a single C\&C server shuts down the botnet partially resulting in less damage to victims.

\begin{wrapfigure}{l}{0.55\textwidth}
  \vspace{-6mm}
    \centering
    \includegraphics[width=0.55\textwidth]{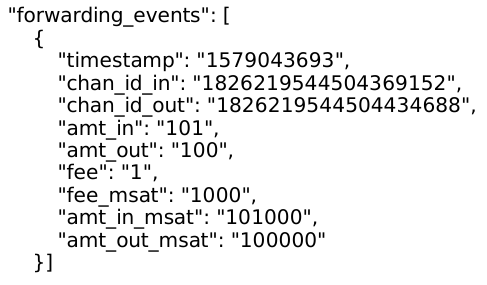}
     \vspace{-6mm}
    \caption{The payment forwarding information stored on Node A's local database in JSON format as the output of the command \textit{lncli fwdinghistory}.}
    \label{fig:forwarding}
      \vspace{-9mm}
\end{wrapfigure}

\noindent $\bullet$ \textit{Payment Flow Timing Analysis for Detecting the Botmaster:} 
As explained in Section \ref{sec:onionrouting}, the intermediary nodes in a payment path do not know the origin of the payment; therefore they cannot distinguish between the botmaster and a regular forwarding node on the payment path unless the payment path just consists of 1-hop \cite{beres2019cryptoeconomic}. In our tests, we observed that our payments took 4 hops to reach C\&C servers. Therefore, payment analysis for such multiple hops is a challenge. However, it can help increase our chances to detect the botmaster. 
\begin{wrapfigure}{r}{0.55\textwidth}
    \centering
   \vspace{-8mm}
    \includegraphics[width=0.5\textwidth]{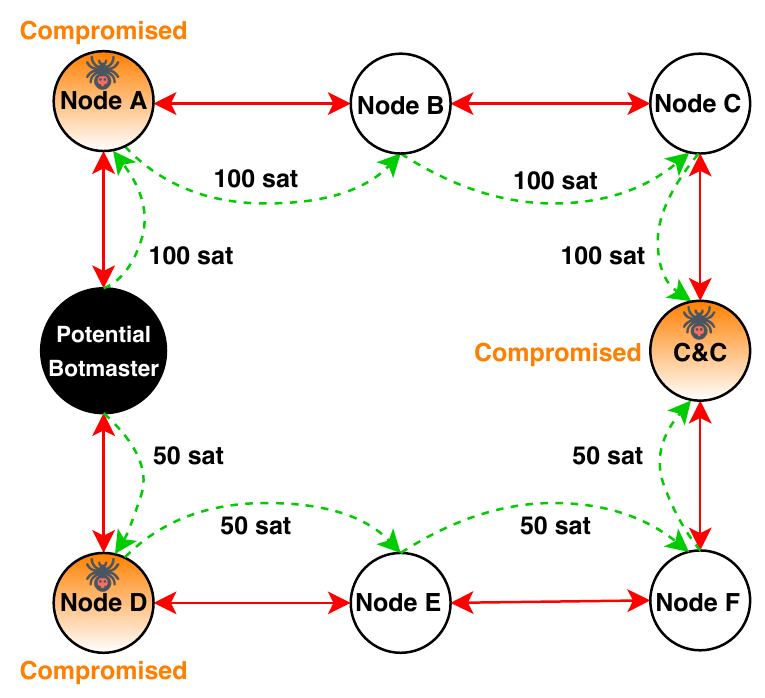}
    \caption{The payments that are forwarded by Node A and Node D are monitored by an observer and the C\&C server is compromised. Red arrows show the payment channels between the nodes and the green arrows show the flow of the payment.}
    \label{fig:trafficanalysis}
   \vspace{-7mm}
\end{wrapfigure}
To further investigate this attack scenario, a topology of 8 nodes was created on Bitcoin Testnet as shown in Fig. \ref{fig:trafficanalysis}. We assume that Node A, Node D and the C\&C server are compromised and thus we monitored their payments. In this setup, a 100 satoshi payment was sent from the botmaster to the C\&C server through hops Node A, Node B, and Node C and the payment was monitored at Node A. By monitoring the node, we got the payment forwarding information shown in Fig. \ref{fig:forwarding}.

In the same way, another 50 satoshi payment was sent from the botmaster to the same C\&C server following hops Node D, Node E, and Node F and the payment was monitored at Node D. Similar payment forwarding information is obtained at node D. Here, particularly important information for us is the timestamp of the payment, and the \textit{chan\_id\_in} and the \textit{chan\_id\_out} arguments which represent the ID of the payment channels that carry the payment in and out from Node A. We can query these channel IDs to learn the public keys of the nodes at both ends of the channel by running \textit{lncli getchaninfo chan\_id}. Obtained LN public keys at Node A, in this case, belong to potential botmaster and Node B. In the same way, LN public keys of potential botmaster and Node E is obtained at Node D. After the payment is observed at Node A, payment with the same amount was observed at the C\&C server. We now correlated these two payments (i.e., timing analysis) and suspected that the sender to Node A (or D) can be a potential botmaster. 
Obviously, there is no guarantee for this (e.g., imagine a different topology where real botmaster is 2 more hops away). We need to collect more data from many compromised nodes and continue this analysis for a long time. To increase the chances, well-connected LN nodes could be requested to cooperate in case of law enforcement investigation to share the timing of payments passing from them.

\noindent $\bullet$ \textit{Poisoning Attack:} 
Another effective way to counter the botmaster is through message poisoning. Basically, once a C\&C server is compromised, its public keys will be known. Using these public keys one can send payments to C\&C servers to corrupt the messages sent by the botmaster at the right time. There is currently no authentication mechanism that can be used by the botmaster without being exposed to prevent this issue. Recall that the commands are encoded in a series of payments and when a different payment is sent during a command transmission, it will corrupt the syntax and thus eventually there will not be any impact. The right time will be decided by listening to the payments and packets arriving at the C\&C server. The disadvantage of this, however, is that one needs to pay for those payments. 
Nonetheless, this can be an effective way to continue engaging with the botmaster for detection purposes rather than just shutting down the C\&C server while rendering any attack impossible. 

\noindent $\bullet$ \textit{Analysis of On-chain Transactions:} Another countermeasure could be through analyzing the on-chain funding transfers of C\&C servers (i.e., channel creation transactions stored on blockchain). For such forensic analysis, the Bitcoin addresses of the C\&C servers should be known. As with many other real-life botnets, botmasters generally use Bitcoin mixers to hide the source of the Bitcoins. Usage of such mixers makes it very hard to follow the real source of the Bitcoins since the transactions are mixed between the users using the mixer service. Even though the chances of finding the identity of the botmaster through this analysis is low, it can provide some useful information to law enforcement.

\section{Related Work}
\label{sec:relatedwork}
Botnets have been around for a long time and there have been even surveys classifying them \cite{bailey2009survey,Grizzard}. While early botnets used IRC, later botnets focused on P2P C\&C for resiliency. 
Furthermore, Tor has also been used for a botnet C\&C but it is shown that botnet activity over Tor can be exposed due to the recognizable patterns in the C\&C communication \cite{casenove2014botnet}. Our proposed LNBot falls under covert botnets which became popular much later. As an example, Nagaraja et al. proposed Stegobot, a covert botnet using social media networks as a command and control channel~\cite{Nagaraja}. Some work has been done by Natarajan et al. to detect Stegobot~\cite{detectStegobot2}. Pantic et al. proposed a covert botnet command and control using Twitter~\cite{Pantic}. Tsiatsikas et al. proposed SDP-Based Covert Channel for Botnet Communication~\cite{SDP}. Calhoun et al. presented a MAC layer covert channel based on WiFi~\cite{calhoun2012802}. 

Recent covert botnets started to utilize Blockchain although these are very few. 
For instance, Roffel et al.~\cite{Roffel} came up with the idea of controlling a computer worm using the Bitcoin blockchain. \cite{Resiliency} discusses how botnet resiliency can be enhanced using private blockchains. Pirozzi et al. presented the use of blockchain as a command and control center for a new generation botnet~\cite{Botchain}. Similarly, ChainChannels \cite{chainchannels} utilizes Bitcoin to disseminate C\&C messages to the bots. 
These works are different from our architecture as they suffer from the issues of high latency and public announcement of commands. There are also Unblockable Chains~\cite{UnblockableChains}, and BOTRACT~\cite{BOTRACT}, which are Ethereum-based botnet command and control infrastructures that suffer from anonymity issues since the commands are publicly recorded on the blockchain. Baden et al. \cite{whisper} proposed a botnet C\&C scheme utilizing Ethereum's Whisper messaging. However, it is still possible to blacklist the topics used by the botmaster. Additionally, there is not a proof of concept implementation of the proposed approach yet, therefore it is unknown if the botnet can successfully be deployed or not.

The closest work to ours are ZombieCoin \cite{ali2015zombiecoin} and Bitcoin Testnet botnet \cite{dazaleveraging}. ZombieCoin uses Bitcoin transaction spreading mechanism as the C\&C communication infrastructure. In this study, the botmaster announces the commands to the bots in terms of legitimate Bitcoin transactions on the Bitcoin network. Then, any legitimate Bitcoin nodes that receive these transactions check the correctness of the input address, the digital signature, and in\&out Bitcoin amounts of the transaction. The bots extract the concealed commands from these transactions. However, this scheme has several drawbacks: First, the authors assumed that the bots identify related transactions from the botmaster's Bitcoin address, which Bitcoin miners can blacklist. Second, as in the case of other blockchain-based botnets, because all transactions are publicly announced, it leaves a public record about the botnet activity. To resolve this problem, in a further study they also proposed to employ subliminal channels \cite{ali2018zombiecoin} to cover the botmaster. However, subliminal channels require a lot of resources to calculate required signatures which is computationally expensive and not practical to use on a large scale.  

Bitcoin Testnet botnet is a recently proposed botnet \cite{dazaleveraging}, where Bitcoin Testnet is utilized for controlling the botnet. Even though their C\&C communication is encrypted, non-standard Bitcoin transactions used for communication exposes the botnet activity. Once the botnet is detected, the messages coming from the botmaster can be prevented from spreading, consequently stopping the botnet activity. Additionally, it is possible for Bitcoin developers to reset the current Bitcoin Testnet (i.e., Testnet3) and create a new Bitcoin testnet (e.g., Testnet4) to stop the botnet completely. 

In contrast, our work is based on legitimate LN payments and does not require any additional computation to hide the commands. Also, these commands are not announced publicly. Moreover, LNBot offers very unique advantage for a botnet that does not contain any direct relation with C\&C. This means even C\&C itself is not aware of the botmaster due to LN's anonymous multi-hop structure. As a result, LNBot does not carry any mentioned disadvantages through its two-layer hybrid architecture and provides  ultra scalability and high anonymity compared to others.

\section{Conclusion}
\label{sec:conclusion}
LN has been formed as a new payment network to address the drawbacks of Bitcoin transactions in terms of time and cost. In addition to relationship anonymity, LN significantly reduces fees by performing off-chain transactions. This provides a perfect opportunity for covert communications as no transactions are recorded in the blockchain. Therefore, in this paper, we proposed a new covert hybrid botnet by utilizing the LN payment network formed for Bitcoin operations. The idea was to control the C\&C servers through messages that are sent in the form of payments through the LN. 
The proof-of-concept implementation of this architecture indicated that LNBot can be successfully created and commands for attacks can be sent to C\&C servers through LN with negligible costs yet with very high anonymity. To minimize LNBot's impact, we offered several countermeasures that include the possibility of searching for the botmaster. 

%
%
%
\bibliographystyle{splncs04}
\bibliography{references}

\begin{thebibliography}{10}
\providecommand{\url}[1]{\texttt{#1}}
\providecommand{\urlprefix}{URL }
\providecommand{\doi}[1]{https://doi.org/#1}

\bibitem{LN-size}
1ml.com: Lightning network search and analysis engine (2019),
  \url{https://1ml.com/}

\bibitem{ali2015zombiecoin}
Ali, S.T., McCorry, P., Lee, P.H.J., Hao, F.: Zombiecoin: powering
  next-generation botnets with bitcoin. In: International Conference on
  Financial Cryptography and Data Security. pp. 34--48. Springer (2015)

\bibitem{ali2018zombiecoin}
Ali, S.T., McCorry, P., Lee, P.H.J., Hao, F.: Zombiecoin 2.0: managing
  next-generation botnets using bitcoin. International Journal of Information
  Security  \textbf{17}(4),  411--422 (2018)

\bibitem{whisper}
Baden, M., Torres, C.F., Pontiveros, B.B.F., State, R.: Whispering botnet
  command and control instructions. In: 2019 Crypto Valley Conference on
  Blockchain Technology (CVCBT). pp. 77--81. IEEE (2019)

\bibitem{bailey2009survey}
Bailey, M., Cooke, E., Jahanian, F., Xu, Y., Karir, M.: A survey of botnet
  technology and defenses. In: Cybersecurity Applications \& Technology
  Conference For Homeland Security, CATCH'09. pp. 299--304. IEEE (2009)

\bibitem{beres2019cryptoeconomic}
B{\'e}res, F., Seres, I.A., Bencz{\'u}r, A.A.: A cryptoeconomic traffic
  analysis of bitcoins lightning network. arXiv preprint arXiv:1911.09432
  (2019)

\bibitem{calhoun2012802}
Calhoun~Jr, T.E., Cao, X., Li, Y., Beyah, R.: An 802.11 mac layer covert
  channel. Wireless Communications and Mobile Computing  \textbf{12}(5),
  393--405 (2012)

\bibitem{casenove2014botnet}
Casenove, M., Miraglia, A.: Botnet over tor: The illusion of hiding. In: 2014
  6th International Conference On Cyber Conflict (CyCon 2014). pp. 273--282.
  IEEE (2014)

\bibitem{Sphinx}
Danezis, G., Goldberg, I.: Sphinx: A compact and provably secure mix format.
  In: 2009 30th IEEE Symposium on Security and Privacy. pp. 269--282. IEEE
  (2009)

\bibitem{dazaleveraging}
Franzoni, F., Abellan, I., Daza, V.: Leveraging bitcoin testnet for
  bidirectional botnet command and control systems. In: Financial Cryptography
  and Data Security 2020

\bibitem{chainchannels}
Frkat, D., Annessi, R., Zseby, T.: Chainchannels: Private botnet communication
  over public blockchains. In: IEEE ITHINGS-GREENCOM-CPSCOM-SMARTDATA 2018. pp.
  1244--1252. IEEE (2018)

\bibitem{Grizzard}
Grizzard, J.B., Sharma, V., Nunnery, C., Kang, B.B., Dagon, D.: Peer-to-peer
  botnets: Overview and case study. HotBots  \textbf{7}, ~1--1 (2007)

\bibitem{huffman1952method}
Huffman, D.A.: A method for the construction of minimum-redundancy codes.
  Proceedings of the IRE  \textbf{40}(9),  1098--1101 (1952)

\bibitem{onionrouting2}
Labs, L.: Bolt \#4: Onion routing protocol (2019),
  \url{https://github.com/lightningnetwork/lightning-rfc/blob/master/04-onion-routing.md}

\bibitem{LN}
Labs, L.: Lightning network daemon (2019), \url{https://lightning.engineering}

\bibitem{gRPC}
Labs, L.: Lnd grpc api reference (2019), \url{https://api.lightning.community/}

\bibitem{samplelndconf}
Labs, L.: Sample lnd.conf (2019),
  \url{https://github.com/lightningnetwork/lnd/blob/master/sample-lnd.conf}

\bibitem{BOTRACT}
Malaika, M.: Botract (2017),
  \url{https://sector.ca/wp-content/uploads/presentations17/Majid-Malaika-Botract\_SecTor.pdf}

\bibitem{Nagaraja}
Nagaraja, S., Houmansadr, A., Piyawongwisal, P., Singh, V., Agarwal, P.,
  Borisov, N.: Stegobot: a covert social network botnet. In: International
  Workshop on Information Hiding. pp. 299--313. Springer (2011)

\bibitem{detectStegobot2}
Natarajan, V., Sheen, S., Anitha, R.: Multilevel analysis to detect covert
  social botnet in multimedia social networks. The Computer Journal
  \textbf{58}(4),  679--687 (2015)

\bibitem{ollmann2009botnet}
Ollmann, G.: Botnet communication topologies. Retrieved September  \textbf{30},
  ~2009 (2009)

\bibitem{sphinxsend}
Osuntokun, O.: New draft sphinx send mode for spontaneous payments (2019),
  \url{https://github.com/lightningnetwork/lnd/pull/2455}

\bibitem{Pantic}
Pantic, N., Husain, M.I.: Covert botnet command and control using twitter. In:
  Proceedings of the 31st annual computer security applications conference. pp.
  171--180. ACM (2015)

\bibitem{pass2015micropayments}
Pass, R., et~al.: Micropayments for decentralized currencies. In: Proceedings
  of the 22nd ACM SIGSAC Conference on Computer and Communications Security.
  pp. 207--218. ACM (2015)

\bibitem{Botchain}
Pirozzi, A., Paganini, P.: Experts presented botchain, the first fully
  functional botnet built upon the bitcoin protocol (2018),
  \url{https://securityaffairs.co/wordpress/77395/malware/botchain-botnet-bitcoin-protocol.html}

\bibitem{poon2015bitcoin}
Poon, J., Dryja, T.: The bitcoin lightning network: Scalable off-chain instant
  payments (2015), \url{https://lightning.network/lightning-network-paper.pdf}

\bibitem{Roffel}
Roffel, D., Garrett, C.: A novel approach for computer worm control using
  decentralized data structures (2014)

\bibitem{SILVA2013378}
Silva, S.S., Silva, R.M., Pinto, R.C., Salles, R.M.: Botnets: A survey.
  Computer Networks  \textbf{57}(2),  378--403 (2013)

\bibitem{Resiliency}
Sweeny, J.: Botnet resiliency via private blockchains (2017),
  \url{https://www.sans.org/reading-room/whitepapers/covert/botnet-resiliency-private-blockchains-38050}

\bibitem{SDP}
Tsiatsikas, Z., Anagnostopoulos, M., Kambourakis, G., Lambrou, S., Geneiatakis,
  D.: Hidden in plain sight. sdp-based covert channel for botnet communication.
  In: International Conference on Trust and Privacy in Digital Business. pp.
  48--59. Springer (2015)

\bibitem{Wang}
Wang, P., Wu, L., Aslam, B., Zou, C.C.: A systematic study on peer-to-peer
  botnets. In: 2009 Proceedings of 18th International Conference on Computer
  Communications and Networks. pp.~1--8. IEEE (2009)

\bibitem{UnblockableChains}
Zohar, O.: Unblockable chains (2018),
  \url{https://github.com/platdrag/UnblockableChains}

\end{thebibliography}
%






\end{document}